\documentclass[aps,prl,twocolumn,showpacs,preprintnumbers,amsmath,amssymb,superscriptaddress]{revtex4}%

\usepackage{graphicx}
\usepackage{dcolumn}
\usepackage{bm}
\usepackage{color}

\begin{document}

\preprint{PREPRINT (\today)}

\newpage
\title{Interplay of the electronic and lattice degrees of freedom in A$_{1-x}$Fe$_{2-y}$Se$_{2}$ superconductors under pressure}

\author{M.~Bendele}
\email{markus.bendele@physik.uzh.ch}
\affiliation{Dipartimento di Fisica, Universit\`{a}~di Roma ``La Sapienza'', P.le Aldo Moro 2, 00185 Roma, Italy}
\affiliation{Rome International Center for Materials Science Superstripes (RICMaSS), Via dei Sabelli 119A, 00185 Roma, Italy}

\author{C. Marini}
\affiliation{European Synchrotron Radiation Facility, BP220, 38043 Grenoble Cedex, France}

\author{B. Joseph}
\affiliation{Dipartimento di Fisica, Universit\`{a}~di Roma ``La Sapienza'', P.le Aldo Moro 2, 00185 Roma, Italy}
\affiliation{Rome International Center for Materials Science Superstripes (RICMaSS), Via dei Sabelli 119A, 00185 Roma, Italy}

\author{G. M. Pierantozzi}
\affiliation{Dipartimento di Fisica, Universit\`{a}~di Roma ``La Sapienza'', P.le Aldo Moro 2, 00185 Roma, Italy}

\author{A. S. Caporale}
\affiliation{Dipartimento di Fisica, Universit\`{a}~di Roma ``La Sapienza'', P.le Aldo Moro 2, 00185 Roma, Italy}
\affiliation{Rome International Center for Materials Science Superstripes (RICMaSS), Via dei Sabelli 119A, 00185 Roma, Italy}

\author{A. Bianconi}
\affiliation{Rome International Center for Materials Science Superstripes (RICMaSS), Via dei Sabelli 119A, 00185 Roma, Italy}

\author{E. Pomjakushina}
\affiliation{Laboratory for Developments and Methods, Paul Scherrer Institut, 5232 Villigen PSI, Switzerland}

\author{K.~Conder}
\affiliation{Laboratory for Developments and Methods, Paul Scherrer Institut, 5232 Villigen PSI, Switzerland}

\author{A. Krzton-Maziopa}
\affiliation{Faculty of Chemistry, Warsaw University of Technology, 00-664 Warsaw, Poland}
\affiliation{Laboratory for Developments and Methods, Paul Scherrer Institut, 5232 Villigen PSI, Switzerland}

\author{ T.  Irifune}
\affiliation{Geodynamics Research Center, Ehime University, 2-5 Bunkyo-cho, Matsuyama
790-8577, Japan}

\author{T. Shinmei}
\affiliation{Geodynamics Research Center, Ehime University, 2-5 Bunkyo-cho, Matsuyama
790-8577, Japan}

\author{S. Pascarelli}
\affiliation{European Synchrotron Radiation Facility, BP220, 38043 Grenoble Cedex, France}

\author{P. Dore}
\affiliation{CNR-SPIN and Dipartimento di Fisica, Universit\`{a}~di Roma ``La Sapienza'', P.le Aldo Moro 2, 00185 Roma, Italy}

\author{N. L. Saini}
\affiliation{Dipartimento di Fisica, Universit\`{a}~di Roma ``La Sapienza'', P.le Aldo Moro 2, 00185 Roma, Italy}

\author{P. Postorino}
\affiliation{CNR-IOM and Dipartimento di Fisica, Universit\`{a}~di Roma ``La Sapienza'', P.le Aldo Moro 2, 00185 Roma, Italy}

\begin{abstract}
The local structure and electronic properties of Rb$_{1-x}$Fe$_{2-y}$Se$_2$ are investigated by means of site selective polarized x-ray absorption spectroscopy at the iron and selenium K-edges as a function of pressure. A combination of dispersive geometry and novel nanodiamond anvil pressure-cell has permitted to reveal a step-like decrease in the Fe-Se bond distance at $p\simeq11$ GPa. The position of the Fe K-edge pre-peak, which is directly related to the position of the chemical potential, remains nearly constant until $\sim6$ GPa, followed by an increase until $p\simeq 11$ GPa. Here, as in the local structure, a step-like decrease of the chemical potential is seen. Thus, the present results provide compelling evidence that the origin of the reemerging superconductivity in $A_{1-x}$Fe$_{2-y}$Se$_2$ in vicinity of a quantum critical transition is caused mainly by the changes in the electronic structure. 
\end{abstract}


\maketitle

Pressure plays an important role in the newly discovered Fe-based materials for tuning the superconducting and magnetic properties owing to their extreme sensitivity to even slight crystallographic distortions \cite{Johnston_2010}. After the discovery of superconductivity in the LaFeAsO$_{1-x}$F$_{x}$ \cite{Kamihara2008} a strong dependence of the superconducting transition temperature $T_{\rm c}$ with the ionic radius of the rare-earth atom was observed \cite{Johnston_2010}. 
Such a remarkable chemical pressure effect has been ascribed to the specific position of the pnictogen/chalcogen (Pn/Ch) atoms above the Fe-plane \cite{Kuroki_2010}.
In this context, the binary superconducting compound FeSe got considerable attention due to its structural simplicity and its similarity in the electronic structure \cite{Hsu_2008}. Furthermore, the system shows extreme sensitivity to pressure since both chemical pressure induced by substitution of Se by Te or S as well as hydrostatic pressure lead to a substantial increase of $T_{\rm c}$ \cite{Medvedev_Nat_09,Bendele_2011,KhasanovFeSeTe}. 
Recently, a new route to increase $T_{\rm c}$ was found by intercalating alkali metals between the FeSe layers leading to $A_{1-x}$Fe$_{2-y}$Se$_{2}$ ($A$ = alkali metal) \cite{Guo10,Krzton11,Li11a}. Interestingly, all these newly discovered materials show a similar $T_{\rm c}\sim 30$ K regardless the size of the intercalated alkali metal atoms. 
The  $A_{1-x}$Fe$_{2-y}$Se$_{2}$ system show a phase separation scenario between an insulating magnetic majority phase and thin metallic stripes which, below $T_{\rm c}$, undergo the superconducting transition \cite{Bendele13,Li11b,Ricci11b,Chen_PRX}. 
It was argued, that during cooling the compressive strain between those two phases leads to such high values of $T_{\rm c}$ comparable to the ones observed in high pressure studies of FeSe indicating an intimate role of the lattice degrees of freedom \cite{Iadecola_12}. Very recently the application of hydrostatic pressure on the A$_{1-x}$Fe$_{2-y}$Se$_{2}$ system lead to a peculiar $T-p$ phase diagram \cite{Sun12}: The value of $T_{\rm c}$ remains almost constant up to $\sim5$ GPa. As the pressure is further increased, $T_{\rm c}$ starts to decrease and vanishes around 9 GPa. The system does not exhibit a superconducting phase from $\sim9$ to about 11 GPa, where the reemergence of superconductivity is observed. It is worth to notice that the $T_{\rm c}$ of the high pressure phase is higher (48 K) than that observed in the low pressure regime (30 K). A quantum critical transition (QCT) close to a structural transition has been hypothesized to separate the two superconducting phases \cite{Guo12}.

In this study the pressure evolution of the electronic and local structures of superconducting Rb$_{1-x}$Fe$_{2-y}$Se$_{2}$ single crystals is investigated under hydrostatic pressures up to $p\sim$ 15 GPa at room temperature (RT) by means of polarized dispersive x-ray absorption spectroscopy at the Fe and Se K-edges. Such a study under pressure is the cleanest way to probe simultaneously the changes in the electronic and lattice structures where in general three important parameters can be obtained: i) the site specific local structure from the partial atomic pair distribution by extended x-ray absorption fine structure (EXAFS); ii) higher orders of the local atomic distribution via x-ray absorption near edge structure (XANES); iii) the position of the chemical potential using the energy shift of the absorption pre-edge feature. Using this local probe, a transition from a disordered to a locally more ordered state at $\sim11$ GPa is identified. 
The electronic subsystem, on the other hand, seems to show a more complex pressure dependence that is mostly driven by the underlying quantum critical point. In addition, it seems to govern the local structural ordering.

The sample was prepared following the procedure described in \cite{Krzton11,Weyeneth12}. Two membrane driven opposing-plate diamond anvil cells (DAC) equipped with 300 $\mu$m culet I-A diamonds for the Fe K-edge, and state of the art 400 $\mu$m culet nano-diamonds \cite{Irifune_03,Ishimatsu_12} for the Se K-edge measurements were used to pressurize the Rb$_{1-x}$Fe$_{2-y}$Se$_{2}$ with KCl as a pressure transmitting medium. The gaskets were made of a 250 $\mu$m thick rhenium foil with a sample chamber of $\sim$130 $\mu$m diameter and 40-50 $\mu$m height. Pressure was measured in-situ exploiting the standard ruby fluorescence technique.
Fe and Se K-edge dispersive x-ray absorption spectroscopy measurements were performed at the energy dispersive EXAFS beamline ID24 of the ESRF, Grenoble \cite{ID-24-1,ID-24-2}. The x-ray source consisted of three undulators of which the gaps were adjusted to tune the maximum of the first and second harmonic at the energy of the Fe K-edge (7112 eV) and Se K-edge (12658 eV), respectively. The beam was focused horizontally by a curved polychromator (Si 111) crystal in Bragg geometry and vertically with a bent Si mirror at a glancing angle of 4 and 2.5 mrad with respect to the direct beam resulting in a beam size at the sample of 3$\times$3 and 15$\times$15 $\mu$m$^2$ for the Fe and Se K-edges, respectively. The XANES spectra were recorded in transmission mode using a FreLon CCD camera detector and calibrated using reference samples.

\begin{figure}[t]
\includegraphics[width=\linewidth]{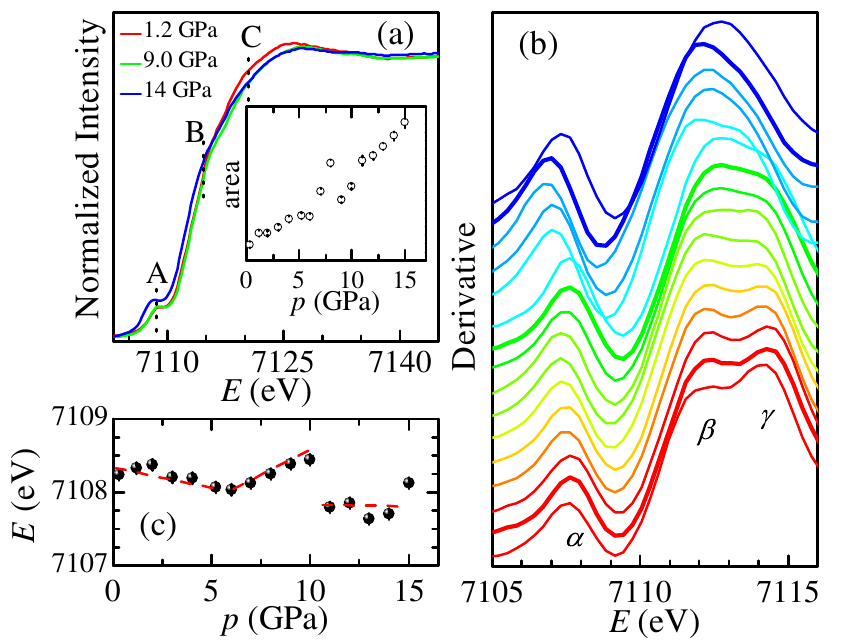}
\caption{(Color online) (a) Normalized Fe K-edge XANES of Rb$_{1-x}$Fe$_{2-y}$Se$_{2}$ at three pressures at room temperature. The pre-edge peak A and the two near-edge features B and C are indicated with dotted lines. The inset shows the pressure evolution of the pre-peak area after subtraction of the background, representing the DOS close to $E_{\rm F}$. (b) The first derivative of the Fe K-edge XANES for pressures ranging from $p\simeq0.3$ to 15 GPa in $\simeq1$ GPa steps. The thicker lines are the corresponding ones shown in (a) with the same color code. (c) Pressure dependence of the inflection point ($\alpha$) representing the chemical potential. The dashed lines are a guide to the eyes.}
\label{Fig_XANES_Fe}
\end{figure}


Normalized Fe K-edge XANES spectra are shown for three pressures in Fig. \ref{Fig_XANES_Fe} (a). The ambient pressure Fe K-edge XANES spectrum appears very similar to those observed in the FeSe system \cite{JosephJPCM,Chen11}. With increasing pressure clear changes in the near edge features are observed. The Fe K-edge is governed mainly by the 1s $\rightarrow \epsilon$p transition with a strong pre-edge feature having substantial contribution of the direct 1s $\rightarrow$ 3d quadrupole transitions to the unoccupied Fe 3d states hybridized with the Se 4p states \cite{JosephJPCM}. With pressure a clear change of the position of the pre-edge peak [labeled as A in Fig. \ref{Fig_XANES_Fe} (a)] is observed. The pressure evolution can readily be seen from the peak $\alpha$ in the derivative spectrum shown Fig. \ref{Fig_XANES_Fe} (b). 
The near-edge features at B and C as indicated in Fig. \ref{Fig_XANES_Fe} (a) reveal a reversal of the spectral weight with increasing pressure. These changes can be appreciated better from the features $\beta$ and $\gamma$ in the derivative spectrum [Fig. \ref{Fig_XANES_Fe} (b)]. At pressures below $p\lesssim6$ GPa feature C is more intense as compared to B. Above $p\simeq 7$ GPa, however, feature B becomes more dominant and above $\sim11$ GPa the weight of feature C is found to be substantially decreased. These changes are a clear indication of the local lattice rearrangement with pressure. In fact, diffraction studies showed a transition from Fe-vacancy ordered ($I4/m$) to disordered ($I4/mmm$) state at this pressure \cite{Guo12}.

Since the pre-edge peak provides information about the chemical potential and the hybridization between the Fe 3d and Se 4p orbitals, that are responsible for the bands close to the Fermi level $E_{\rm F}$ \cite{Liu12b,Chen12}, a detailed investigation is undertaken. From the maximum of the derivative representing the inflection point, the pressure evolution of the chemical potential is extracted [Fig. \ref{Fig_XANES_Fe} (c)]. Until $p\simeq6$ GPa it shows a tendency to decrease, followed by an increase with increasing pressure. At $p\simeq11$ GPa it shows a sudden decrease to lower energies. Since the density of states (DOS) close to $E_{\rm F}$ is related to the pre-peak area, its pressure dependence can be more quantitatively evaluated by looking at the integrated intensity obtained after proper background subtraction. As shown in the inset to Fig. \ref{Fig_XANES_Fe} (a) it continuously increases with increasing pressure. Thus, the compression of the lattice indicates an increase of the DOS showing that the system is becoming more metallic. 

\begin{figure}[t]
\includegraphics[width=\linewidth]{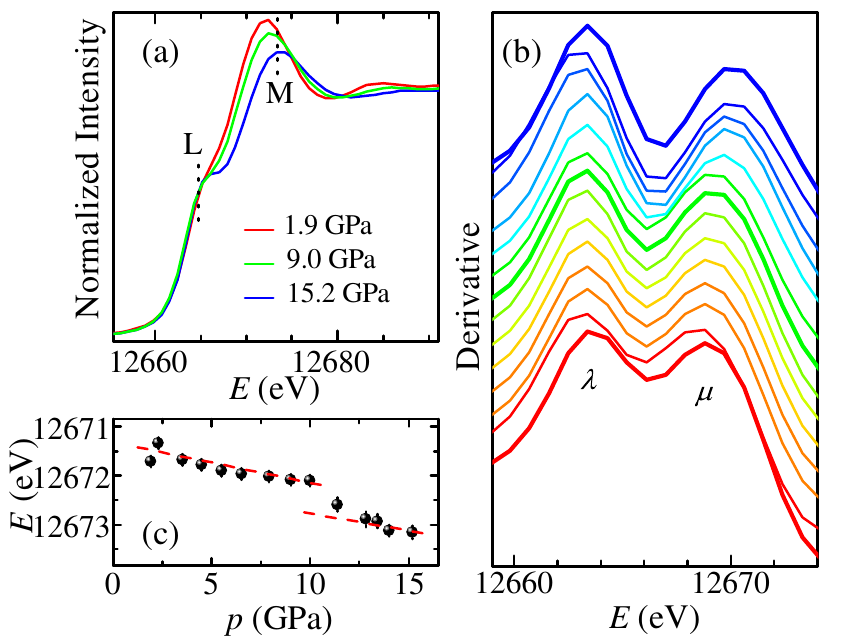}
\caption{(Color online) (a) Normalized Se K-edge XANES of Rb$_{1-x}$Fe$_{2-y}$Se$_{2}$ at room temperature for three pressures. (b) The first derivative of the Se K-edge for all pressures ranging from ambient pressure to 15 GPa in $\simeq1$ GPa steps. The thicker lines are the corresponding ones shown in (a) with the same color code. (c) Pressure dependence of the maximum of the multiple scattering peak M. The dashed line in (c) is a guide to the eyes.}
\label{Fig_XANES_Se}
\end{figure}

Normalized Se K-edge XANES spectra of Rb$_{1-x}$Fe$_{2-y}$Se$_{2}$ are shown in Fig. \ref{Fig_XANES_Se}(a). It mainly consists of two features: a sharp peak L around 12665 eV due to the direct 1s $\rightarrow$ 4p dipole transition and a broad hump M about 8 eV above the edge which is mainly governed by the multiple scattering of the photo-electron with its neighbors and closely related to the local atomic arrangement \cite{JosephJPCM}. Already in Fig. \ref{Fig_XANES_Se}(a) it can be seen that the position of the sharp peak is hardly affected by pressure. On the contrary the feature arising from the multiple scattering is largely influenced by pressure. In order to investigate the evolution in detail, the first derivatives are considered [Fig. \ref{Fig_XANES_Se} (b)]. As already visible in the raw spectra, a change can be observed mainly in the multiple scattering feature [see $\mu$ in Fig. \ref{Fig_XANES_Se}]. The pressure dependence of the latter feature M is shown in Fig. \ref{Fig_XANES_Se}(c). The position continuously increases with increasing pressure until 10 GPa due to the compression of the lattice. At $\simeq 11$ GPa, however, an abrupt change of the compressibility is observed. Above it appears to decrease again continuously with a similar slope as compared to low pressures. 


\begin{figure}[t]
\includegraphics[width=\linewidth]{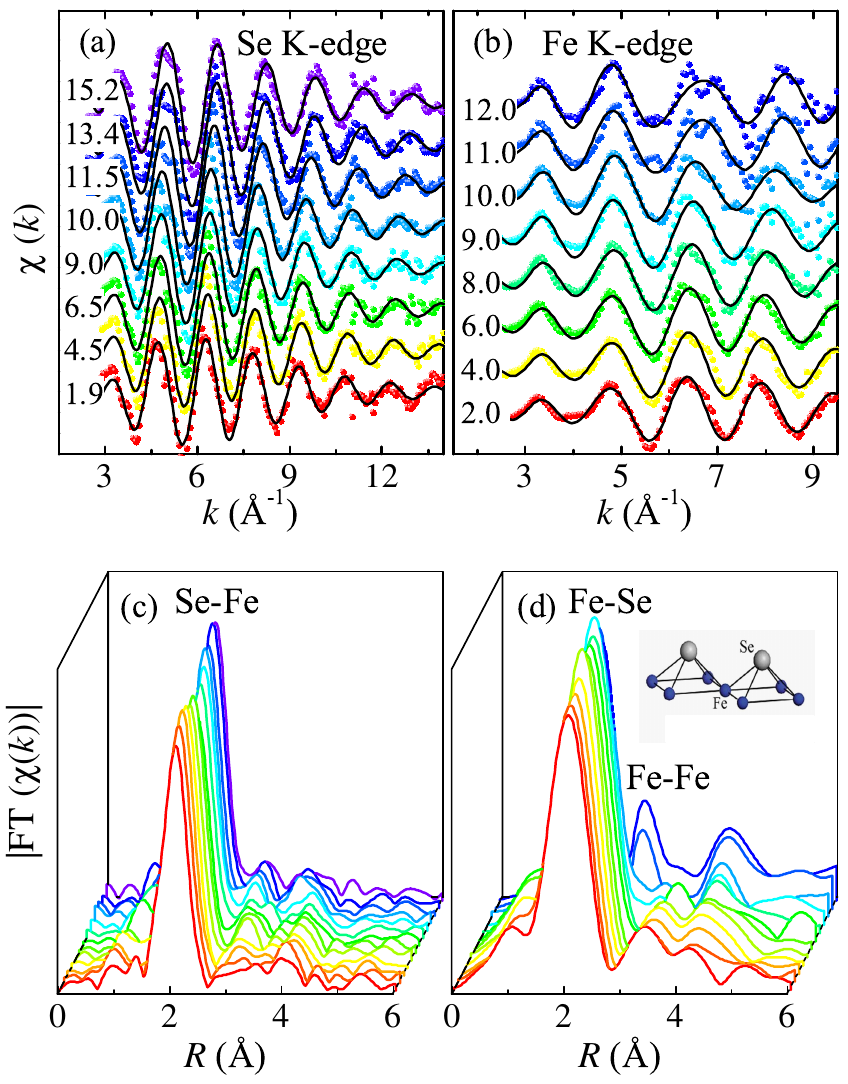}
\caption{(Color online) The EXAFS oscillations of Rb$_{1-x}$Fe$_{2-y}$Se$_2$ at room temperature extracted from the x-ray absorption spectra for selected pressures of (a) Se and (b) Fe K-edges. The data are shifted vertically for clarity in presentation. The measured data points are shown as symbols whereas the first shell modeling results involving the Fe-Se bond distance for the Se K-edge and the Fe-Se and Fe-Fe bond distances for the Fe K-edge are shown as solid black lines. Fourier transform (FT) magnitudes of the EXAFS oscillations corresponding to Se K-edge (c) and Fe K-edge (d). }
\label{Fig_EXAFS}
\end{figure}

In order to probe the lattice response of the observed reorganization of the electronic structure in Rb$_{1-x}$Fe$_{2-y}$Se$_2$, the local atomic distribution around the Se and Fe along the direction of x-ray beam polarization is probed by means of polarized EXAFS. The obtained spectra for selected pressures up to $p\simeq15$ and $\simeq12$ GPa from the Se and Fe K-edges at RT are presented in Figs. \ref{Fig_EXAFS}(a) and (b), respectively. Note the larger $k$-ranges achieved from the nano- compared to the I-A diamonds. In both edges a systematic shift of the oscillations towards higher $k$ is observed for increasing pressure consistent with the decreasing of the bond distances. 
Interestingly, the Fe K-edge EXAFS shows a significant change between 10 and 11 GPa most visible in the oscillation at $k\simeq 6.5$ \AA$^{-1}$ that seems to broaden significantly and split [see Fig. \ref{Fig_EXAFS}(b)]. 

Earlier ambient pressure Se K-edge EXAFS studies on the A$_{1-x}$Fe$_{2-y}$Se$_2$ system have underlined the presence of large local disorder in this system \cite{Iadecola_12,Tyson12}. The atomic distributions above the first-shell were found to be almost completely suppressed, i.e. the nearest neighbors Rb and Se are not visible. The low pressure data shown in Fig. \ref{Fig_EXAFS} (a) are consistent with these studies. Quantitative bond distributions can be obtained from the analysis of the data. The first shell modelling involving near neighbor atoms are presented as solid lines in Fig. \ref{Fig_EXAFS} (a).
As already seen in the EXAFS oscillations, the peak representing the Fe-Se distance in the Fourier transform (FT) shifts to lower $R$ corresponding to a decrease in the lattice parameter with increasing pressure [Fig. \ref{Fig_EXAFS} (c)]. Interestingly, with increasing pressure, the distribution of higher shells seem to become more pronounced. 

The Fe K-edge EXAFS, on the other hand, is dominated by the Fe-Se and Fe-Fe bond-distances, which appear at distances of $d_{\rm Fe-Se}\sim 2.4$ and $d_{\rm Fe-Fe}\sim 2.7$ \AA, respectively [Fig. \ref{Fig_EXAFS} (d)]. However, the Fe-Fe bond is hardly visible in the EXAFS even though all nearest neighbor Fe ions are aligned in the $ab$-plane, i.e. parallel to the beam polarization [inset to Fig. \ref{Fig_EXAFS} (d)]. This is due to the large local disorder at the Fe site \cite{Iadecola_12,Tyson12}. By monitoring the pressure evolution of the FT of the Fe K-edge EXAFS, the decrease of the Fe-Se distance seen already in the Se K-edge is confirmed. However, as already mentioned above, a significant change is observed between 10 and 11 GPa. In the FT a new peak appears at the position where contributions from the Fe-Fe bonds are supposed to be present ($R\sim2.6$ \AA). This sudden appearance of the Fe-Fe bond contributions indicates a substantial local structural change at pressures above $\simeq11$ GPa. 

\begin{figure}[t]
\includegraphics[width=\linewidth]{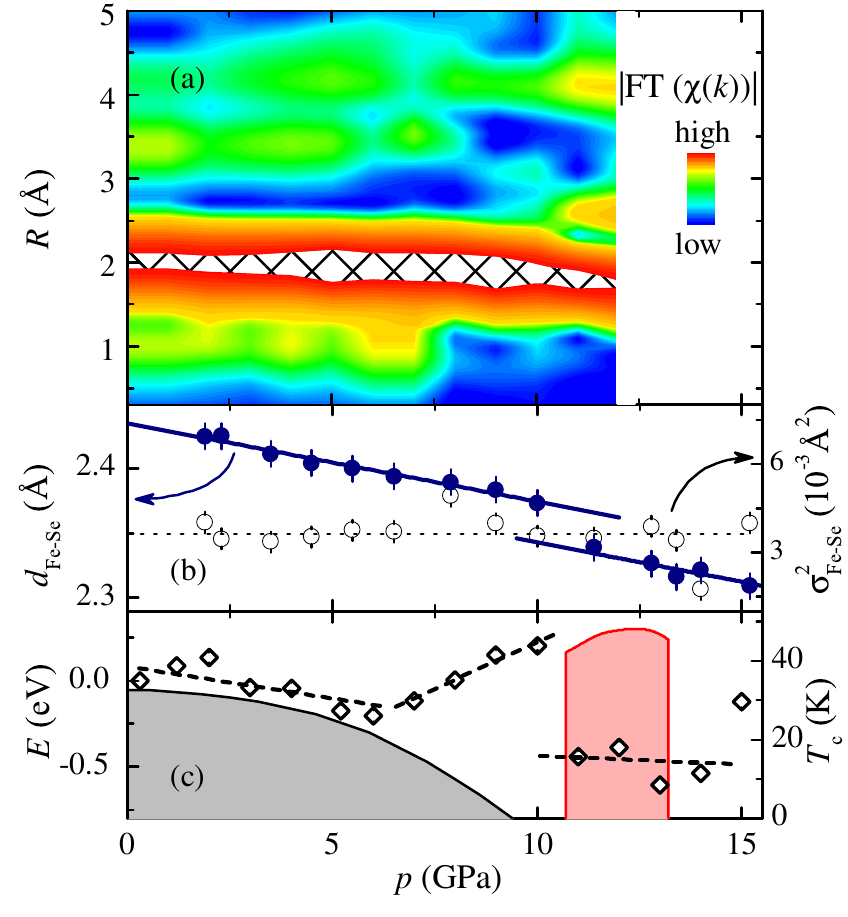}
\caption{(Color online) (a) Pressure dependence of the FT of the Fe K-edge EXAFS of Rb$_{1-x}$Fe$_{2-y}$Se$_2$ at room temperature in a 2D contour plot. (b) The Fe-Se distance (left axis) and the corresponding Debye-Waller factor $\sigma^2_{\rm Fe-Se}$ (right axis) derived from the first shell analysis of the Se K-edge EXAFS. (c) The pressure dependence of the chemical potential with respect to ambient pressure (left axis) and the schematic pressure dependence of the superconducting transition temperature $T_{\rm c}$ adapted from Ref. \cite{Sun12} (right axis).}
\label{Fig_Dist}
\end{figure}

This is illustrated in detail in the contour plot of the FT of the Fe K-edge EXAFS presented in Fig. \ref{Fig_Dist} (a). With increasing pressure the Fe-Se distance is decreasing (white line). Although highly damped in intensity, the contour plot also provides the opportunity to monitor the higher shells ($R\sim4$ \AA). Here the changes up to $\sim7$ GPa seem to be rather insignificant, compared to which there seems a significant change between $7-11$ GPa. Above $p\simeq11$ GPa the Fe-Fe peak ($R\sim2.5$ \AA) and the higher shell contributions ($R\sim4$ \AA) are clearly visible. The same behavior of an increased local order above $p\sim11$ GPa is seen from the Se K-edge measurements [Fig. \ref{Fig_EXAFS} (c)]. At the same pressure the Fe-Se distance $d_{\rm Fe-Se}$, obtained from the first shell analysis of the Se K-edge EXAFS measurements, reveals a step like decrease [Fig. \ref{Fig_Dist} (b)]. At lower and higher pressures $d_{\rm Fe-Se}$ is continuously decreasing with increasing pressure in agreement with the observation of a continuous decrease of the unit cell volume \cite{Svitlyk11,Ksenofontov12,Guo12}. 
This depicts the observed transition from a tetragonal vacancy ordered phase at low pressures to a high pressure phase \cite{Guo12}. 

In Fig. \ref{Fig_Dist} (c) the schematic pressure dependence of $T_{\rm c}$ adapted from Ref. \cite{Sun12} is shown alongside the pressure dependence of the chemical potential obtained from the Fe K-edge measurements with respect to ambient pressure. At pressures up to $\sim5$ GPa where $T_{\rm c}$ remains nearly constant, the chemical potential is slightly decreasing. In this pressure range, the bond Debye Waller factor ($\sigma^2_{\rm Fe-Se}$) obtained from the Se K-edge measurements [right axis Fig. \ref{Fig_Dist} (b)] remains nearly constant. In the pressure range where the superconducting transition temperature decreases, the chemical potential is increasing with increasing pressure [Fig. \ref{Fig_Dist} (c)]. Interestingly, here $\sigma^2_{\rm Fe-Se}$ shows a larger value giving an indication for a possible topological transition of the electronic structure. As mentioned earlier, in this pressure range the system seems to be more disordered since in the FT the higher shells are suppressed. As soon as the system crosses the suggested QCT from a disordered to a locally more ordered phase, superconductivity reemerges, the chemical potential shows a sudden decrease, and the magnetic ordering vanishes \cite{Guo12,Ksenofontov12}. After crossing of the second superconducting phase, the chemical potential seems to increase again, which is supported by yet another anomalous change of $\sigma^2_{\rm Fe-Se}$ in this pressure range. Since before the QCT the chemical potential shows a strong change, one might speculate that the changes in the local lattice are driven by changes in the electronic structure.

In conclusion, the present study unravels a lattice rearrangement to a locally more ordered state at $p\simeq 11$ GPa in Rb$_{1-x}$Fe$_{2-y}$Se$_2$, where in addition a step like decrease of the Fe-Se bond distance occurs. This transition seems to be driven electroncally since the chemical potential shows an increase in the region $p\sim 7 - 11$ GPa followed by an unusual step-like decrease at $p\simeq11$ GPa. These results provide an indication that an underlying QCT is associated with the reemerging superconductivity in alkali-metal intercalated iron chalcogenides.



The authors thank I. Kantor (ID-24) for his help during the data collection. M.B. acknowledges the financial support of the Swiss National Science Foundation (grant number PBZHP2\_143495).


\begin{thebibliography}{99}

\bibitem{Johnston_2010} D. C. Johnston, Adv. Phys. \textbf{59}, 803 (2010).

\bibitem{Kamihara2008} Y. Kamihara, T. Watanabe, M. Hirano and H. Hosono, J. Am. Chem. Soc. \textbf{130}, 3296 (2008).

\bibitem{Kuroki_2010} K. Kuroki, H. Usui, S. Onari, R. Arita, and H. Aoki, Phys. Rev. B \textbf{79}, 224511 (2009).


\bibitem{Hsu_2008} F.~C.~Hsu, J.~Y.~Luo, K.-W.~Yeh, T.-K.~Chen, T.-W.~Huang, P.-M.~Wu, Y.-C.~Lee, Y.-L.~Huang, Y.-Y.~Chu, D.-C.~Yan, and M.-K.~Wu, Proc. Natl Acad. Sci. USA \textbf{105}, 14262 (2008).

\bibitem{Bendele_2011} M.~Bendele, A.~Amato, K.~Conder, M.~Elender, H.~Keller, H.-H.~Klauss, H.~Luetkens, E.~Pomjakushina, A.~Raselli, and R.~Khasanov, Phys. Rev. Lett. \textbf{104}, 087003 (2010).

\bibitem{KhasanovFeSeTe} R.~Khasanov, M.~Bendele, A.~Amato, P.~Babkevich, A.~T.~Boothroyd, A.~Cervellino, K.~Conder, S.~N.~Gvasaliya, H.~Keller, H.-H.~Klauss, H.~Luetkens, V.~Pomjakushin, E.~Pomjakushina, and B.~Roessli, Phys. Rev. B \textbf{80}, 140511(R) (2009).

\bibitem{Medvedev_Nat_09} S.~Medvedev, T.M.~McQueen, I.A.~Troyan,  T.~Palasyuk, M.I.~Eremets,  R.J.~Cava,  S.~Naghavi, F.~Casper, V.~Ksenofontov, G.~Wortmann, and C.~Felser, 
Nature~Mater. \textbf{8}, 630 (2009).


\bibitem{Guo10} J. Guo, S. Jin, G.Wang, S.Wang, K. Zhu, T. Zhou, M. He, and X. Chen, 
Phys. Rev. B \textbf{82}, 180520(R) (2010).

\bibitem{Krzton11} A. Krzton-Maziopa, Z. Shermadini, E. Pomjakushina, V. Pomjakushin, M. Bendele, A. Amato, R. Khasanov, H. Luetkens, and K. Conder, 
J. Phys.: Condens. Matter \textbf{23}, 052203 (2011).


\bibitem{Li11a} C.-H. Li, B. Shen, F. Han, X. Zhu, and H.-H. Wen,
Phys. Rev. B \textbf{83}, 184521 (2011).

\bibitem{Li11b} W. Li, H. Ding, P. Deng, K. Chang, C. Song, K. He, L, Wang, X. Ma, J.-P. Hu, X. Chen, and Q.K. Xue, 
Nature Phys. \textbf{8}, 126 (2011).

\bibitem{Ricci11b} A. Ricci, N. Poccia, B. Joseph, G. Arrighetti, L. Barba, J. Plaisier, G. Campi, Y. Mizuguchi, H. Takeya, Y. Takano, N. L. Saini, and A. Bianconi, 
Supercond. Sci. Technol. \textbf{24}, 082002 (2011).

\bibitem{Bendele13} S. C. Speller, T. B. Britton, G. M. Hughes, A. Krzton-Maziopa, E. Pomjakushina, K. Conder, A. Boothroyd, and C. R. Grovenor, Supercond. Sci. Tenchnol. \textbf{25}, 984023 (2012).

\bibitem{Chen_PRX} F. Chen, M. Xu, Q. Q. Ge, Y. Zhang, Z. R. Ye, L. X. Yang, Juan Jiang, B. P. Xie, R. C. Che, M. Zhang, A. F. Wang, X. H. Chen, D. W. Shen, J. P. Hu, and D. L. Feng, 
Phys. Rev. X \textbf{1}, 021020 (2011).


\bibitem{Iadecola_12} A. Iadecola, B. Joseph, L. Simonelli, A. Puri, Y. Mizuguchi, H. Takeya, Y. Takano, and N. L. Saini, J. Phys.: Condens. Matter \textbf{24}, 115701 (2012).

\bibitem{Sun12} L. Sun, X.-J. Chen, J. Guo, P. Gao, Q.-Z. Huang, H. Wang, M. Fang, X. Chen, G. Chen, Q. Wu, C. Zhang, D. Gu, X. Dong, L. Wang, K. Yang, A. Li, X. Dai, H.-k. Mao, and Z. Zhao, Nature \textbf{483}, 67 (2012).

\bibitem{Guo12} J. Guo, X.-J. Chen, J. Dai, C. Zhang, J. Guo, X. Chen, Q. Wu, D. Gu, P. Gao, L. Yang, K. Yang, X. Dai, H.-k. Mao, L. Sun, and Z. Zhao, Phys. Rev. Lett. \textbf{108}, 197001 (2012).

\bibitem{Weyeneth12} S. Weyeneth, M. Bendele, F. von Rohr, P. Dluzewski, R. Puzniak, A. Krzton-Maziopa, S. Bosma, Z. Guguchia, R. Khasanov, Z. Shermadini, A. Amato, E. Pomjakushina, K. Conder, A. Schilling, and H. Keller, Phys. Rev. B \textbf{86}, 134530 (2012).

\bibitem{Irifune_03} T.  Irifune, A. Kurio, S. Sakamoto, T. Inoue, and H. Sumiya, Nature \textbf{421}, 599 (2003).

\bibitem{Ishimatsu_12} N. Ishimatsu, K. Matsumoto, H. Maruyama, N. Kawamura, M. Mizumaki, H. Sumiya, and T. Irifune, J. Synchrotron Rad. \textbf{19},  768 (2012).



%

\bibitem{ID-24-1} S. Pascarelli, O. Mathon, M. Munoz, T. Mairs, and J. Susini, J. Synchrotron Rad. \textbf{13}, 351 (2006).

\bibitem{ID-24-2} S. Pascarelli and O. Mathon, Phys. Chem. Chem. Phys. \textbf{12}, 5535 (2010). 
%

\bibitem{JosephJPCM} B. Joseph, A. Iadecola, L. Simonelli, Y. Mizuguchi, Y. Takano, T. Mizokawa, and N.L. Saini, J. Phys.: Condens. Matter \textbf{22}, 485702 (2010).

\bibitem{Chen11}C. L. Chen, S. M. Rao, C. L. Dong, J. L. Chen, T. W. Huang, B. H. Mok, M. C. Ling, W. C. Wang, C. L. Chang, T. S. Chan, J. F. Lee, J.-H. Guo, and M. K. Wu, EPL \textbf{93}, 47003 (2011).

\bibitem{Ksenofontov12} Vadim Ksenofontov, Sergey A. Medvedev, Leslie M. Schoop, Gerhard Wortmann, Taras Palasyuk, Vladimir Tsurkan, Joachim Deisenhofer, Alois Loidl, and Claudia Felser, Phys. Rev. B \textbf{85}, 214519 (2012).

\bibitem{Svitlyk11} V. Svitlyk, D. Chernyshov, E. Pomjakushina, A. Krzton-Maziopa, K. Conder, V. Pomjakushin, and V. Dmitriev, Inorg. Chem. \textbf{50}, 10703 (2011).

\bibitem{Tyson12} T. A. Tyson, T. Yu, S. J. Han, M. Croft, G. D. Gu, I. K. Dimitrov, and Q. Li, Phys. Rev. B \textbf{85}, 024504 (2012).
\bibitem{Liu12b} Z.-H. Liu, P. Richard, N. Xu, G. Xu, Y. Li, X.-C. Fang, L.-L. Jia, G.-F. Chen, D.-M. Wang, J.-B. He, T. Qian, J.-P. Hu, H. Ding, and S.-C. Wang, 
Phys. Rev. Lett. \textbf{109}, 037003 (2012).

\bibitem{Chen12} Chen Fei, Q. Q. Ge, M. Xu, Y. Zhang, X.P. Shen, W. Li, M. Matsunami, S.-i. Kimua, J.P. Hu, and D. L. Feng,
Chin. Sci. Bull. \textbf{57}, 3829 (2012).

\end{thebibliography}
\end{document}